\documentclass[aps,twocolumn,groupedaddress,showpacs,amssymb,prl,floatfix,
preprintnumbers]{revtex4}
\usepackage{graphicx,color}

\newcommand{\nbco}       {${\rm Nd} {\rm Ba}_{2} {\rm Cu}_{3} {\rm O}_{6+y}$}
\newcommand{\nbcnzo}     {NdBa$_2$\-(Cu,\-Ni,\-Zn)$_3$O$_{6+y}$}

\newcommand{\ybcox}      {${\rm Y} {\rm Ba}_{2} {\rm Cu}_{3} {\rm O}_{6+y}$}

\newcommand{\cudf}       {$^{63,65}$Cu}
\newcommand{\cuot}       {CuO$_2$}
\newcommand{\etal}       {{\it et~al}.}

\newcommand{\lsxco}      {${\rm La}_{1.85} {\rm Sr}_{0.15} {\rm Cu O_4}$}

\newcommand{\ybcoacht}   {$\rm YBa_2 Cu_4 O_{8}$}

\definecolor{orange}{rgb}{0.7,0.6,0.1}

\begin{document}

\thispagestyle{myheadings}

\title{Cu NQR Wipeout Effect versus Charge Pseudogap in Zn/Ni Doped
${\rm\bf Nd} {\rm\bf Ba}_{\bf 2} {\rm\bf Cu}_{\bf3} {\rm\bf
O}_{\bf 6+y}$}

\author{H.-J. Grafe$^{1}$, F. Hammerath$^{1}$, A. Vyalikh$^{1}$, G. Urbanik$^1$,
V. Kataev$^{1}$, Th. Wolf$^{2}$, G. Khaliullin$^{3}$, and B.
B\"uchner$^1$} \affiliation{$^1$IFW Dresden, Institut f\"ur Festk\"orperforschung, P.O. Box 270116, D-01171 Dresden, Germany\\
$^2$Forschungszentrum Karlsruhe, Institut f\"ur Festk\"orperphysik, D-76021 Karlsruhe, Germany\\
$^{3}$Max-Planck-Institut f\"ur Festk\"orperforschung,
Heisenbergstrasse 1, D-70569 Stuttgart, Germany}


\date{\today}

\begin{abstract}
We report \cudf\ NQR measurements on slightly underdoped \nbco\ single
crystals heavily doped by Ni and Zn impurities. Owing to the impurity
doping superconductivity is fully suppressed in both cases. The Ni
strongly enhances magnetic correlations and induces a wipeout of the NQR
signal comparable to that found in stripe ordered lanthanum cuprates. In
contrast, the magnetism is suppressed in the Zn doped sample where no
wipeout effect is observed and the nuclear spin relaxation rate is
reduced. Our findings are in a striking correspondence with the different
impact of Ni and Zn impurities on the charge pseudogap evidenced by recent
optical data, uncovering thereby a close relationship between the magnetic
correlations and pseudogap phenomena.
\end{abstract}


\maketitle

Magnetic (Ni) and non-magnetic (Zn) substitution for copper in
superconducting cuprates can help to understand the still unveiled
mechanism of high temperature superconductivity (HTSC) by studying
the influence of these impurities on the electronic and magnetic
properties of the CuO$_2$ planes. Adding a few percent of Ni or Zn
is sufficient to suppresses superconductivity in
La$_{2-x}$Sr$_x$CuO$_4$ (LSCO) \cite{Xiao90}. Ni and Zn doping
affects also the unusual normal state electronic properties of
HTSCs such as the pseudogap (PG) state \cite{Timusk99}. In
particular, observation of the magnetic moments induced by a
nominally spinless Zn impurity in underdoped cuprates
\cite{Finkelstein90,Alloul91,Sidis00,Julien00,gvmw2,gvmw3,itoh2}
has been one of the most spectacular manifestations of nontrivial
correlations operating in the PG state. Recently it was found
\cite{bernhard} that higher amounts of Ni and Zn can be
incorporated in \nbco\ (NBCO), and thereby superconductivity is
fully suppressed even at optimal doping. Measurements of the
optical conductivity that probes the \textit{charge} excitations
in these compounds show that Ni and Zn have a profoundly different
impact on the PG \cite{bernhard}. Large Zn-doping suppresses the
PG, whereas Ni enhances its energy scale. Such a drastic
difference suggests that the nonmagnetic and magnetic doping may
have opposite effect on the \textit{spin} dynamics and
antiferromagnetic (AF) correlations in the PG regime.

To elucidate the influence of the impurity doping on the spin
dynamics we have performed \cudf\ nuclear quadrupole resonance
(NQR) measurements of single crystals of Ni and Zn doped NBCO.
Since in addition to CuO$_2$ planes responsible for HTSC the
crystal structure comprises CuO chains we have estimated the
distribution of the Ni and Zn ions between these two structural
units by analyzing the Cu NQR lineshape. The most striking result
of our study is a Ni induced wipeout of the Cu NQR signal at low
temperatures similar to that observed in stripe ordered rare earth
(RE) co-doped lanthanum cuprates
\cite{nick,hunt,singer,hunt2,barbara}. The wipeout is caused by a
continuously slowing down of the Cu electronic spin fluctuations.
In contrast the Zn doping does not induce any wipeout albeit a
significantly larger concentration of the Zn in the planes
compared to Ni. Here the intensity as well as the linewidth and
shape of the spectra is unchanged down to 4.2\,K. Also the spin
lattice relaxation is suppressed by the Zn in the whole
temperature range. This difference in the spin dynamics is in a
striking correspondence with the different charge dynamics
measured by optical conductivity. Doping by Ni leads to a slowing
of electronic spin fluctuations and a glassy magnetic order and
enhances the energy scale of the charge PG. On the other hand Zn
doping does not induce such a quasistatic magnetic order and
thereby a wipeout of the NQR signal, but suppresses the charge PG
and the spin lattice relaxation rate. Thus our data point out an
intimate interplay between the spin- and charge excitations in the
PG state of the cuprates and support theoretical predictions of
the different effect of magnetic and non-magnetic impurities on
the properties of the strongly correlated CuO$_2$ planes in HTSCs.

We used high quality single crystals of \nbcnzo\ with 11.5\,$\%$ Ni,
5\,$\%$ Zn, and 9.5\,$\%$ Zn, and a superconducting NBCO sample without
impurities as a reference. All samples are slightly underdoped with
$y\,\approx\,0.9$. For details of the sample preparation see Ref.
\cite{bernhard}. Midpoint $T_c$ is 83\,$\pm$4\,K in the pure sample and
26\,$\pm$5\,K in the 5\,$\%$ Zn doped sample.  Note that $T_c$ in
underdoped NBCO is lower than in underdoped  \ybcox\  (YBCO).
Superconductivity is completely suppressed in the samples with 11.5\,$\%$
Ni and 9.5\,$\%$ Zn.
A tuned NMR/NQR circuit with an almost $T$-independent low quality factor
$Q$ similar to that in Ref. \cite{gvmw} was used to minimize the influence
of $Q$ on the spectrum intensity $^{\rm Cu}I_{\rm NQR}$. The intensity was
also corrected for the Boltzmann factor and for spin echo decay $T_2$
\cite{hunt2}. The spin lattice relaxation time, $T_1$ was measured by
inversion recovery of the longitudinal magnetization $M(t)$ which was
fitted to the expression
$M(t)$\,=\,$M(0)$\,$\cdot$\,$(1-2\exp(-(3t/T_1)^{\lambda})$, where
$\lambda\leq 1$.

\begin{figure}
\begin{center}
 \includegraphics[width=70mm,clip]{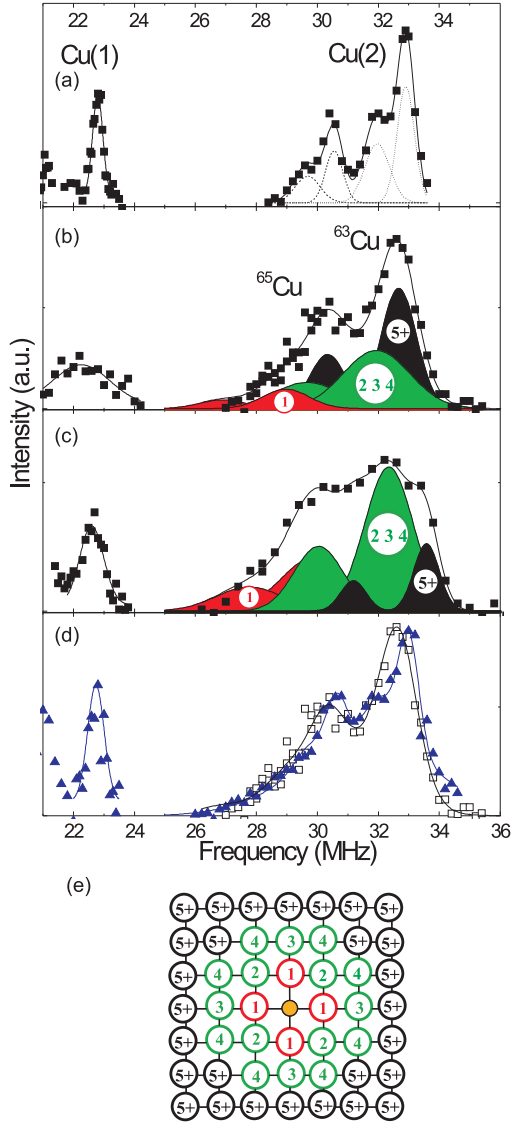}
 \caption{\cudf\ NQR spectra of NBCO for different dopings. Cu(1) denotes the signal
  from the chains, and Cu(2) the planar spectra. All spectra were taken at $T$\,=\,200\,K.
  Panel (a): pure NBCO , (b) 11.5\,$\%$ Ni doped sample, (c) 9.5\,$\%$ Zn doped
  sample, (d) comparison of the spectrum of the 11.5\,$\%$ Ni doped sample ($\square$) with
  the spectrum of the 5\,$\%$ Zn doped sample ($\textcolor{blue}{\blacktriangle}$), (e) Schematic
  picture of the Cu sites around an impurity site ($\textcolor{orange}{\bullet}$). The fits
  are explained in the text.}
  \label{spectra}
\end{center}
\end{figure}

Fig.~\ref{spectra} shows \cudf\ NQR spectra of the chains and planes of
NBCO for different dopings at $T$\,=\,200\,K.
The resonance frequency $\nu_{\rm NQR}$ of the planar Cu(2) in
pure NBCO is blue-shifted in comparison to YBCO by
$\sim$\,1.5\,MHz owing to the larger ionic radius of the Nd
compared to the Y \cite{itoh}, but the linewidth and shape
corresponds well to that in YBCO \cite{itoh2,vega,keren}. In
contrast to the RE elements, Zn and Ni doping on the Cu site
affects $\nu_{\rm NQR}$ only slightly, but changes significantly
the linewidth and shape. Thereby Ni broadens the chain signal
stronger than Zn, and Zn broadens the planar signal stronger than
Ni. This difference may occur since Ni preferably occupies the
chain Cu site in YBCO whereas Zn substitutes mainly for the Cu in
the planes \cite{bernhard}. To estimate the impurity content in
the CuO$_2$ planes we have fitted the spectra by three lines for
each Cu isotope using a model similar to that used by Itoh \etal\
\cite{itoh2}. The black shaded line in Fig.~\ref{spectra}
corresponds to those Cu sites that are 5-th and farther nearest
neighbors to the impurity (5+NNs) \cite{comment}. The light grey
(online green) shaded line originates from the Cu sites that are
closer to an impurity, namely the 4NNs to 2NNs. Such a line has
also been found before in Zn doped YBCO and \ybcoacht\
\cite{itoh2,gvmw2,gvmw3,commentx}, but not for the Ni doped
samples, probably due to a smaller Ni content in the planes (see
below). Finally, the dark grey (red online) shaded line
corresponds to the 1NNs Cu sites to the impurity. Such a line has
been suspected by Itoh \etal\ \cite{itoh2}, but could not be
verified in the spectra. It has been assumed that this line as
well as the missing line in case of Ni doping \cite{gvmw2,gvmw3}
are wiped out. These lines are however not wiped out in our
spectra. We have then numerically calculated the number of
particular NN Cu sites around an impurity, and compared them with
the intensity of the respective lines of the fits. From that we
obtain $\sim$\,4\,\%\,Ni and $\sim$\,9.5\,\%\,Zn impurities in the
CuO$_2$ planes. Since there are two CuO$_2$ layers and one chain
layer in the unit cell we calculate for the chains an impurity
content of 9.5\,$\%$ Zn and 26.5\,$\%$ Ni. That the Ni prefers the
chain sites is evident from a much stronger broadening of the
chain spectrum in this case compared to the Zn doped sample.
Moreover, we compare in Fig.\ref{spectra} (d) a spectrum of a
sample with 5\,$\%$ Zn with the spectrum of the Ni doped sample.
Clearly, the planar Cu(2) spectra of both samples are very similar
whereas the chain Cu(1) signal of the Zn sample is only weakly
broadened which evidences a preferential planar site occupancy by
Zn. Despite the controversial discussion about the Cu site
assignment in the CuO$_2$ planes around a defect
\cite{julien,gvmw4,bersier}, our data supports a model that
distinguishes different Cu sites close to an impurity by Cu NQR.
Also the frequency dependence of the relaxation rate $T_1^{-1}$
supports this model (see below).

\begin{figure}
\begin{center}
 \includegraphics[width=72mm,clip]{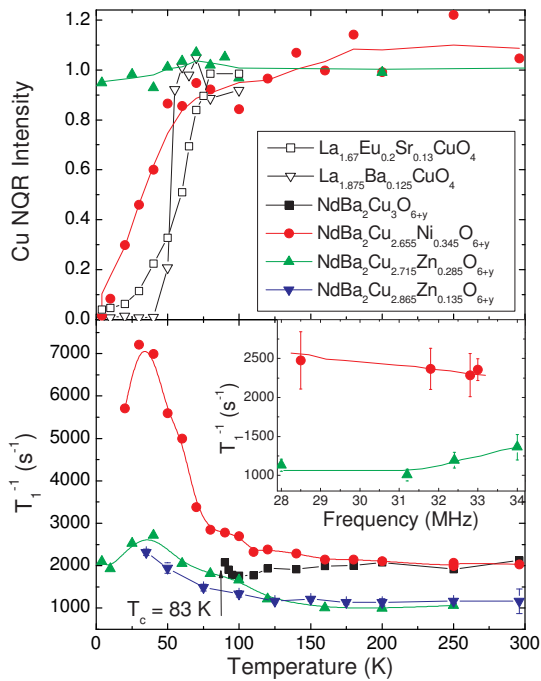}
 \caption{\tolerance=100 \mbox{Upper panel:} $^{\rm Cu}I_{\rm NQR}$ vs. $T$ in NBCO in
 comparison to lanthanum cupra\-tes. \mbox{Lower panel}: $T_1^{-1}(T)$ dependence in NBCO.
 Note its strong enhancement at low $T$s for the Ni, and the suppression in the whole $T$ range
 for both the 5\,$\%$ and 9.5\,$\%$ Zn doped NBCO. Inset: frequency dependence of $T_1^{-1}$
 at 150\,K for the 9.5\,$\%$ Zn and the Ni doped sample. Data at 28 and 28.5\,MHz were multiplied
 by $(^{63}\gamma_{N} / ^{65}\gamma_{N})^2$. Lines are guides to the eye.}
 \label{intens}
\end{center}
\end{figure}

Our most obvious observation that proves the different effect of
Zn and Ni doping on the electronic properties of the CuO$_2$
planes is a wipeout of the Cu NQR signal in the Ni doped sample
below $\sim\,50$\,K, whereas the Cu NQR intensity $^{\rm Cu}I_{\rm
NQR}$ of the Zn doped sample is constant down to the lowest $T$
\cite{comment2}. Also $\nu_{NQR}$, the linewidth and shape of the
spectra of the Zn doped sample exhibit almost no $T$ dependence,
whereas the spectra of the Ni samples change with $T$, especially
in the region where the wipeout sets in. The $^{\rm Cu}I_{\rm
NQR}(T)$ dependences are shown in Fig. \ref{intens} and compared
with those obtained for an oriented powder of ${\rm La}_{1.67}
{\rm Eu}_{0.2} {\rm Sr}_{0.13} {\rm Cu O_4}$ (LESCO) and a single
crystal of ${\rm La}_{1.875} {\rm Ba}_{0.125} {\rm Cu O_4}$
(LBCO). The onset of wipeout in the Ni doped NBCO occurs at a
slightly lower $T$ and develops in a broader temperature range in
comparison to LESCO and LBCO. In lanthanum cuprates the wipeout of
the Cu signal is well known, and has been interpreted as the
stripe order parameter and as evidence for a glassy, inhomogeneous
freezing of Cu spin fluctuations
\cite{hunt,singer,hunt2,nick,barbara}. However, to our knowledge a
complete wipeout of the Cu NQR intensity for an almost optimally
doped RBCO (R= RE or Y) has not been reported before
\cite{comment3}.

In NQR the wipeout occurs when the Cu electronic spin fluctuation
frequency $\tau^{-1}$ decreases with $T$ and approaches
$\nu_{NQR}$. At this point the field $h_0$ at the nuclear site
caused by the fluctuating electronic spins is most efficient in
shortening the nuclear spin relaxation rate $T_1^{-1}= \gamma_N
h_0^2\tau/(1 + 4\pi^2\nu_{NQR}^2\tau^2)$ so that the NQR response
relaxes before it can be measured. Here $\gamma_N$ is the
gyromagnetic ratio. The slowing down of spin fluctuations with
decreasing $T$ in Ni doped NBCO is evidenced by the strong
enhancement of $T_1^{-1}$ (Fig. \ref{intens}) concomitant the
wipeout. Remarkably, Zn doping has an entirely different effect:
$T_1^{-1}$ is suppressed even compared to the pure NBCO at high
$T$ and increases only slightly in the low-$T$ regime. The latter
is probably related to the fluctuating Nd moments whose
susceptibility increases with decreasing $T$ \cite{comment4}. The
frequency dependence of $T_1^{-1}$ (inset of Fig. \ref{intens})
clearly demonstrates the different effect of Ni and Zn on the spin
relaxation and supports the model of the site assignment in the
CuO$_2$ planes: for the Zn doped sample $T_1^{-1}$ decreases with
decreasing the frequency, i.e. by approaching the impurity site
(cf. Fig. \ref{spectra}), whereas it tends to increase in the Ni
doped case. The magnetization decay $M(t)$ is single exponential
($\lambda \approx 1$) at high temperatures for the Ni doped and
the pure NBCO. Remarkably we find $\lambda \leq 1$ for the Zn
doped sample, and below $\sim$70\,K for the Ni doped sample,
implying the distribution of the rates $T_1^{-1}$, similar to the
Zn doped YBCO \cite{gvmw4,walstedt}, and stripe ordered LSCO at
temperatures where the wipeout sets in \cite{nick}. Supporting
evidence for the different spin dynamics in Ni and Zn doped NBCO
comes from $\mu$SR measurements of this compound \cite{bernhard}.
It reveals static magnetic moments on the time scale of $\mu$SR in
case of Ni doping but not for Zn doping. This is also a feature of
the stripe order in lanthanum cuprates \cite{hanshenning}. In
particular, the weakening of the spin correlations by large Zn
doping was found in a $\mu$SR study of LSCO \cite{Adachi04}.

The distinction of NQR data between the Zn- and Ni-substituted
NBCO samples is remarkable. Given that a {\it small} amount of
both Zn and Ni impurities induce magnetic moments and enhance AF
correlations in cuprates
\cite{Finkelstein90,Alloul91,Sidis00,Julien00,gvmw2,gvmw3,itoh2},
their opposite impact on low energy spin dynamics of CuO$_2$
planes that we observe here is highly surprising. Unexpected
suppression of AF correlations by {\it large} Zn doping -- in
contrast to the Ni case -- suggests that a picture of Zn induced
magnetic moments and associated enhancement of AF is no longer
applicable at large Zn-doping. The reason is that while Ni ion
introduces its own local spin $S=1$ coupled antiferromagnetically
to the neighboring Cu spins, the Zn induced magnetic moment has a
completely different microscopic origin
\cite{Khaliullin97,Pepin98,Khaliullin99}. Namely, it can be viewed
as a spatially extended bound state near the Fermi energy,
originating from the many-body response of a correlated Cu-spin
background on a "missing-spin" defect. Because of their highly
nonlocal hence fragile nature, low-energy resonances induced by
different Zn ions should strongly overlap and may eventually
disappear at larger Zn-doping because of the destructive
interference. In this limit, nonmagnetic Zn impurities would
mainly act to suppress low-energy spin collective modes via the
dilution and/or disorder effects \cite{Greven02} amplified at the
presence of charge carriers. Apparently, this is what we see here
as a decrease of NQR relaxation rates. On the contrary, Ni induced
moments are robust because of their local origin, and thus have a
positive impact on AF at all the doping levels \cite{note1}.

It seems that there is a one-to-one correspondence between our
finding that the spectral weight of low-energy magnetic
excitations in heavily Ni (Zn) doped cuprates is enhanced
(suppressed), and that of Ref.~\cite{bernhard} reporting an
increase (decrease) of the optical pseudogap by Ni (Zn) doping. A
straightforward implication of this comparison is that quasistatic
(as probed here by NQR) AF correlations are essential for the
understanding of a celebrated pseudogap phenomenon in cuprates.
Clearly, an enhancement of both PG and AF by Ni doping would be
difficult to rationalize solely in terms of binding of spins into
singlet-pairs, as this would lead to a suppression of the nuclear
spin relaxation rate in contradiction to what is observed.
Therefore, both AF correlations and pairing effects should be
considered on equal footing, which remains a challenge for theory.

One may speculate that similar to the RE doped LSCO
\cite{Tranquada99} the renormalized classical regime with a
reduced spin stiffness can be recovered in the Ni doped NBCO in
spite of a substantial concentration of holes in the planes. In
this scenario the frequency of the spin fluctuations may decrease
exponentially with decreasing temperature causing the wipeout of
the NQR signal, as found in the RE doped LSCO
\cite{nick,hunt,singer} and widely discussed in connection with
the stripe physics, i.e. spatial modulations of the spin density
in the \cuot\ planes. Structural distortions induced by RE
co-doping of LSCO result in the pinning of dynamic stripes and
suppression of superconductivity \cite{Tranquada99}. These
findings in RE doped LSCO are in an apparent similarity with our
data that show a strongly enhanced nuclear relaxation rate
$T_1^{-1}$ (slow electron spin dynamics) in the heavily Ni doped
NBCO. One can speculate therefore on a possible relevance of the
stripe scenario to the physics of NBCO.

Regardless a particular model, our Cu NQR results on the different effect
of the Zn and Ni doping on the spin dynamics in NBCO suggest it as a
generic property of the HTSC cuprates. The slowing down of the electronic
spin fluctuations should have a direct implication on the dynamics in the
charge sector. Indeed, the recovery of the antiferromagnetism owing to the
strong interaction between the Ni impurities enhances the tendency of the
hole localization and thus inhibits the charge dynamics. For this reason
the PG energy scale is pushed up to much higher energies as compared with
the pure NBCO as it has recently been observed \cite{bernhard}.

In summary, our Cu NQR study of the heavily Ni doped NBCO reveals
a complete wipeout of the NQR intensity concomitant with the
strong enhancement of the nuclear relaxation rate $T_1^{-1}$. In a
striking contrast the Zn doping yields a reduction of $T_1^{-1}$
and does not cause a wipeout, although the impurity content of Zn
in the planes is much higher. The data enlighten an entirely
different effect of the magnetic and nonmagnetic doping on the
spin dynamics and correlations in the CuO$_2$ planes that is in a
remarkable correspondence with the different impact of the Ni and
Zn impurities on the charge excitations probed by optical
measurements. In particular, the Cu NQR results provide strong
experimental evidence that magnetic correlations play a vital role
in the charge PG phenomenon in cuprates.

We would like to thank J. Haase and N.J. Curro for helpful
discussions. This work was supported by the Deutsche
Forschungsgemeinschaft, FG 538.


\begin{thebibliography}{99}
\bibitem{comment}
We have neglected in the fit the weak splitting of the resonance
due to the oxygen deficiency that is observed in the pure NBCO
(Fig. \ref{spectra}(a)).
\bibitem{commentx} This line is attributed to 1NN in Ref.~\onlinecite{gvmw2}.
\bibitem{comment3}
A "... significant loss of the Cu NMR intensity ..."  mentioned in
Ref.~\cite{walstedt} for Zn doped YBCO disagrees with our data,
but that partial wipeout was not further quantified.
\bibitem{comment4}
Indeed, in heavily Zn doped RBCO with nonmagnetic Eu$^{3+}$ at the
R site our recent  $T_1^{-1}$ data does not reveal this low-$T$
upturn.
\bibitem{note1} More specifically, Ni impurity can be described in terms
of underscreened Kondo physics, resulting in a net effective spin one-half
which resides on Ni site \cite{Khaliullin99}. These localized spins induce
a low-energy component in the Cu-spin background and may promote a static
AF at sufficiently large Ni doping.
\end{thebibliography}
\end{document}